\documentclass[letterpaper, 10 pt, conference]{ieeeconf}  % Comment this line out if you need a4paper

\IEEEoverridecommandlockouts                              % This command is only needed if 
\usepackage{hyperref}
\usepackage{amsmath,graphicx}
\usepackage{amsfonts}
\usepackage{amsmath}
\usepackage{adjustbox}
\usepackage{bbm}
\usepackage{epstopdf}
\usepackage{fancyhdr,lipsum}
\usepackage{algorithmicx}
\usepackage{algpseudocode}
\usepackage[ruled]{algorithm}
\usepackage{graphicx}
\usepackage{caption}
\usepackage{graphicx,times,amsmath} % Add all your packages here
\usepackage{caption}
\usepackage{subcaption}
\usepackage[rightcaption]{sidecap}
\usepackage{caption}
\usepackage{multirow}
\usepackage{hhline}
\usepackage{amsmath}
\usepackage{epstopdf}
\usepackage{tabularx}
\usepackage{tikz}
\usepackage{tikz}
\usetikzlibrary{matrix}
\usepackage{mathtools, nccmath}
\usepackage{pst-node}
\usepackage{bbm}
\usepackage[absolute,overlay]{textpos}

\usepackage{alphalph}

\title{\LARGE \bf
Hypertension Detection From High-Dimensional Representation of Photoplethysmogram Signals}

\author{Navid Hasanzadeh$^{1}$, Shahrokh Valaee$^{1}$, \textit{Fellow}, \textit{IEEE}, and Hojjat Salehinejad$^{2}$, \textit{Senior Member}, \textit{IEEE} % <-this % stops a space
\thanks{*This work was not supported by any organization.}% <-this % stops a space
\thanks{$^{1}$N. Hasanzadeh and S. Valaee are with the Department of Electrical \& Computer Engineering, University of Toronto, Toronto, ON, Canada. {\tt\small  navid.hasanzadeh@mail.utoronto.ca, valaee@ece.utoronto.ca.}}%        
\thanks{$^{2}$H. Salehinejad is with Kern Center for the Science of Health Care Delivery, Mayo Clinic, Rochester, MN, USA. {\tt\small hojjat@ieee.org.}}%
}

\begin{document}
% \fancyhead[C]{FOR AUTHOR ONLY}
% \toappear{2023 IEEE International Workshop on Machine Learning for Signal Processing, Sept.\ 17--20, 2023, Rome, Italy}
\begin{textblock*}{20cm}(0.6cm,0.4cm) % {block width} (coords) 
   IEEE-EMBS International Conference on Biomedical and Health Informatics (BHI’23), Oct.\ 15--18, 2023, Pittsburgh, Pennsylvania, USA
\end{textblock*}
\maketitle
\thispagestyle{empty}
\pagestyle{empty}

%%%%%%%%%%%%%%%%%%%%%%%%%%%%%%%%%%%%%%%%%%%%%%%%%%%%%%%%%%%%%%%%%%%%
\begin{abstract}

% Hypertension is commonly referred to as ``silent killer", since it can lead to severe health complications without any visible symptoms. Early detection of hypertension is crucial in preventing significant health issues. Although some studies suggest a relationship between blood pressure and certain vital signals, such as Photoplethysmogram (PPG), reliable generalization of the proposed blood pressure estimation methods is not guaranteed yet. This has resulted in some studies doubting the existence of such relationships or considering them weak and limited to heart rate and blood pressure. In this paper, a high-dimensional representation technique based on random convolution kernels is proposed for hypertension detection using PPG signals. The results show this relationship extends beyond heart rate and blood pressure, demonstrating the feasibility of hypertension detection with generalization on unseen subjects. 

Hypertension is commonly referred to as the ``silent killer", since it can lead to severe health complications without any visible symptoms. Early detection of hypertension is crucial in preventing significant health issues. Although some studies suggest a relationship between blood pressure and certain vital signals, such as Photoplethysmogram (PPG), reliable generalization of the proposed blood pressure estimation methods is not yet guaranteed. This lack of certainty has resulted in some studies doubting the existence of such relationships, or considering them weak and limited to heart rate and blood pressure. In this paper, a high-dimensional representation technique based on random convolution kernels is proposed for hypertension detection using PPG signals. The results show that this relationship extends beyond heart rate and blood pressure, demonstrating the feasibility of hypertension detection with generalization. Additionally, the utilized transform using convolution kernels, as an end-to-end time-series feature extractor, outperforms the methods proposed in the previous studies and state-of-the-art deep learning models.

\indent \textit{Clinical relevance}— The findings of this study highlights the feasibility of hypertension detection using PPG signals. This could be useful for the early detection of high blood pressure and reducing the risk of hypertension going unnoticed, particularly using wearable devices such as smartwatches equipped with PPG sensors.
\end{abstract}

%%%%%%%%%%%%%%%%%%%%%%%%%%%%%%%%%%%%%%%%%%%%%%%%%%%%%%%%%%%%%%%%%%%%%%%%%%%%%%%%
\section{Introduction}

Hypertension, or high blood pressure (BP), is a common and dangerous condition that can lead to serious health problems, including heart failure and brain stroke \cite{kalehoff2020story}. It is estimated that 1.28 billion adults aged 30–79 years may have hypertension worldwide~\cite{world2022world}. According to the Centers for Disease Control and Prevention, nearly half of adults in the United States may have hypertension~\cite{HypertensionUSA}. 

People with hypertension are not often aware of it for years. Early detection of hypertension is critical to prevent serious health issues for people at risk. Regular BP checks are recommended for everyone, especially for those at risk, including people with a family history of hypertension, diabetes, or obesity. Early detection allows for early intervention, such as lifestyle changes and medication, which can help to manage hypertension and prevent further complications. However, cuff-based or wrist BP monitoring devices are not available for everyone. These devices are not convenient to use for many people, particularly the elderly. Manual regular BP monitoring is generally inconvenient and requires commitment. These challenges have prompted researchers to seek alternative methods in measuring BP and detecting hypertension \cite{zhao2023emerging}.

Photoplethysmogram (PPG) is a signal collected from an optical sensor which shows fluctuations of the blood volume per heartbeat~\cite{hasanzadeh2023multi}. PPG has recently been investigated as an alternative for continuous BP monitoring without using a cuff. This is particularly of interest as the pattern of an invasive arterial blood pressure (ABP) signal is very similar to the PPG. 
Previous studies have shown that the properties of a PPG signal can indicate various characteristics of the cardiovascular system  \cite{padilla2006assessment, wang2009noninvasive}, such as large artery stiffness index (LASI), systemic vascular resistance (SVR), arterial tone, total peripheral resistance, and pulse wave velocity (PWV). Therefore, by extracting PPG key points and relevant cardiovascular features and applying various machine learning algorithms, BP estimation may be possible \cite{kachuee2016cuffless, aguet2023blood}.

The performance of most PPG-based BP estimation methods has been reported as either very high or very low. Many studies reporting high performance used the UCI cuff-less BP estimation dataset \cite{kachuee2016cuffless}, which comprises $12,000$ PPG and arterial BP signal segments recorded from approximately $1,000$ patients. However, this dataset does not provide any subject identifier (ID) for each PPG signal sample, and the preprocessing steps are not discussed in detail. As a subject may have more than one PPG sample in the dataset, this could lead to data and domain overlap in the training and validation phases. Hence, for methods developed based on this dataset, generalization to a completely unseen subject cannot be guaranteed.

Other PPG datasets typically have a small number of samples \cite{liang2018new}, which limits proper generalization evaluation of machine learning methods for real-world scenarios. This is particularly important in training deep learning models such as recurrent neural networks~\cite{salehinejad2017recent}, where generally a very large number of training samples are required.  
% Majority of the proposed method for hypertension prediction using PPG signals have reported different performance results on different datasets. 
The difference in reported results and the lack of proven generalization on unseen subjects have caused some studies to cast doubt on the existence of any relationship between BP and PPG features \cite{weber2023intensive, bulhoes2022blood}. These studies suggest that the only feature relevant to BP might be the heart rate \cite{mehta2023can}.

% The lack of information about the dataset, as well as the different approaches used for data splitting in previous studies, makes it difficult to assess the generalization performance of the proposed models when applied to completely unseen data, despite reporting very good accuracies. 

% , the various implementations of the same methods could be another reason behind the differences in the reported accuracies.

\begin{figure*}[t]
	\centering
	\subfloat{%
		\includegraphics[width=0.95\linewidth]{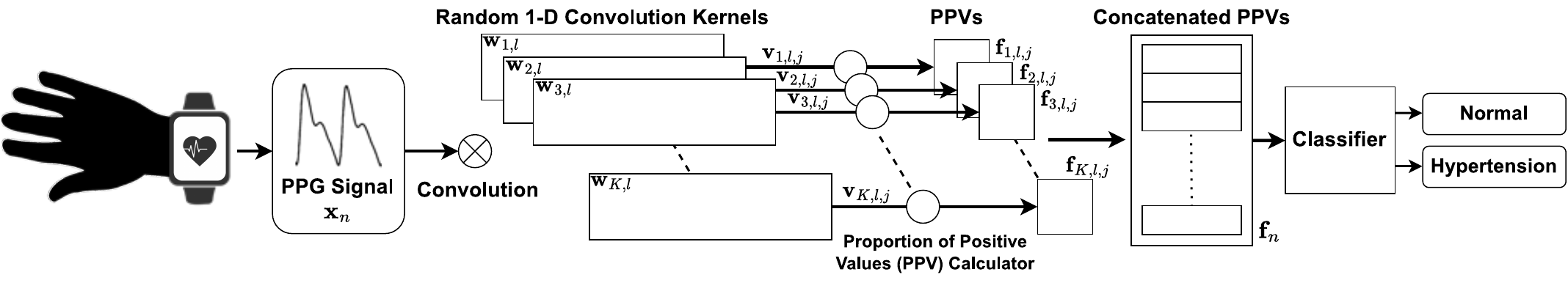}
	}
	\caption{Random convolution kernels for feature extraction and hypertension detection from PPG.}
	\vspace{-4mm}
\label{figure:diagram}
\end{figure*}

This paper addresses the problem of PPG-based BP estimation as a binary hypertension detection task. To this end, the MIMIC-III PPG-BP dataset is used where the train, validation, and test sets are completely separated based on patient IDs \cite{schrumpf2021assessment}. In order to have an end-to-end feature extraction and classification solution, 
an input PPG signal is projected to a high dimensional space using random convolutional kernels transform (ROCKET) \cite{dempster2021MiniROCKET}. The transform maps a time series with any length to a set of temporally-independent features. The extracted features from all the PPG signals are then used to train a classifier.
 Results show a better performance of the proposed method in comparison with manual feature extraction and state-of-the-art deep learning models. The results further support the relationship between PPG properties and hypertension\footnote{Our codes are available online: \url{https://github.com/navidhasanzadeh/Hypertension_PPG}}.

\section{Method}

Feature extraction and classification with random convolution kernels, without training the kernels, is a novel method for time-series representation~\cite{dempster2021MiniROCKET,salehinejad2022s}. This approach has demonstrated a promising performance in many time-series classification tasks such as in electroencephalogram (EEG)~\cite{lundy2021random} and human activity recognition~\cite{salehinejad2022litehar,salehinejad2023joint}. It also has the potential to outperform deep neural networks in many scenarios such as where limited-imbalanced data is available. 

% MiniROCKET is a streamlined version of ROCKET which uses a smaller number of kernels with restrictions.

Figure~\ref{figure:diagram} shows different steps of the proposed method for feature extraction from PPG signals and hypertension detection. Let $\{(\mathbf{x}_1,y_1),...,(\mathbf{x}_N,y_N)\}$ represents a set of $N$ PPG signals where $y_n\in\{0,1\}$ is the data class, with $y=0$ represents normal and $y=1$ representing hypertension.

A set of $K$ 1-dimensional random convolution kernels $(\mathbf{w}_1,...,\mathbf{w}_K)$ are generated where the length of each kernel is $9$ and the weights are selected randomly from $\{-1,2\}$ in such a way that each kernel contains three weights with a value of $2$, and the total sum of the weights in each kernel is zero. Then, a set of dilation factors controls the spread of each kernel over an input PPG signal with length $T$ selected from ${\{\lfloor2^{i\cdot L_{max}/L'}\rfloor|i\in(0,...,L')\}}$ where $L'$ is a constant, $L_{max}=log_2\big((T-1)/(|\mathbf{w}_k|-1)\big)$ and $L$ is the number of constructed dilations,~\cite{dempster2021MiniROCKET,salehinejad2023joint}. 
This provides $K\times L$ different combinations of kernels and dilations as ${\{\mathbf{w}_{k,l}|k\in(1,...,K),l\in(1,...,L)\}}$. 
Each kernel is then convolved with a PPG signal $\mathbf{x}$ as
\begin{equation}
    \mathbf{u}_{k,l} = \mathbf{x}*\mathbf{w}_{k,l},
\end{equation}
for $k\in(1,...,K)$, and $l\in(1,...,L)$. 
Based on the quantiles of the convolution output and for each pair of kernel and dilation $(k,l)$, a set of bias terms $\{b_{k,l,j}|j\in(1,...,J)\}$ is computed. Each bias term shifts the convolution output to generate a new representation as  
\begin{equation}
    \mathbf{v}_{k,l,j}= \mathbf{u}_{k,l} - \mathbf{b}_{k,l,j},
    \label{eq:bias}
\end{equation}
where $\mathbf{b}_{k,l,j} = \underbrace{%
        \begin{pmatrix}
            b_{k,l,j} & \cdots & b_{k,l,j}
        \end{pmatrix}%
     }_{|\mathbf{u}_{k,l}| \text{ times}}$ and $j\in(1,...,J)$, ${k\in(1,...,K)}$, and $l\in(1,...,L)$. The total number of extracted features is a multiple of the number of output features. The output features, called proportion of positive values (PPV), are extracted as
\begin{equation}
    f_{k,l,j}=\frac{1}{|\mathbf{v}_{k,l,j}|}\sum_{i=1}^{|\mathbf{v}_{k,l,j}|}\mathbbm{1}[v_{k,l,j,i}>0],
    \label{eq:ppv}
\end{equation}
for $k\in(1,...,K)$, $l\in(1,...,L)$, and $j\in(1,...,J_{k,l})$ where $J_{k,l}$ is the number of bias terms and $\mathbbm{1}[\cdot]$ is the indicator function. Finally, the extracted features can be represented as $\mathbf{f}=(f_{1},...,f_{D})$ where $D$ is the number of output features. The generated features $\mathbf{f}_n$ for $n\in\{1,...,N\}$ along with the corresponding labels are then used to separately train Ridge regression (RR) and Random Forest (RF) classifiers.

\section{Experiments}

\subsection{Data}

In this study, the PPG signals and corresponding BPs derived from the MIMIC-III dataset \cite{schrumpf2021assessment} are used. The BP values are categorized into normal and hypertension classes based on ESC/ESH guidelines \cite{mancia20142013}. This dataset comprises $3750$ subjects for training and $625$ subjects for testing. The training and test datasets are standardized and divided at subject level to avoid any overlap. There are $1,000,000$ PPG signals for training, $250,000$ samples for validation, and $250,000$ samples for testing. Each PPG sample has a duration of $7$ seconds, and the sampling rate is $125$ Hz. 

% are evaluating models for hypertension detection using the dataset that was introduced in \cite{schrumpf2021assessment}. The dataset comprises PPG signals and corresponding BPs, and it is derived from the PhysioNet MIMIC-III database \cite{johnson2016mimic}. 
% It has been divided into training, validation, and test sets based on subjects to prevent any contamination of the validation and test sets by the training data. The dataset comprises 3750 subjects for training and 625 subjects for validation and testing. As a result, there are 1,000,000 PPG signals for training, 250,000 samples for validation, and 250,000 samples for testing. Each PPG sample has a duration of 7 seconds, and the sampling rate is 125 Hz. 

% In this study, the BP values are categorized into two classes normal and hypertension based on ESC/ESH guidelines. Indeed, subjects with Systolic BP (SBP) values higher than 140 mmHg and/or Diastolic BP (DBP) greater than 90 mmHg are considered as hypertensive subjects and the rest subjects categorized as normal.

% Moreover, in order to study the effect of the size of training set on the models' performance, we also train all the models with $6.25\%$, $12.5\%$, $25\%$, and $50\%$ of the training samples. 

\bgroup
\def\arraystretch{1.4}% 
\begin{table*}[t]
\centering
	\caption{Hypertension detection results by different methods}
			\begin{adjustbox}{width=0.8\textwidth}

	\begin{tabular}{c|c|cc|ccc}		
      \hline

		\hline
		\multirow{2}{*}{Method} &
		\multirow{2}{*}{Classifier} &		
		\multicolumn{2}{c|}{Sensitivity} &	\multicolumn{3}{c}{Weighted Average} \\
		& & Normal & Hypertension & Precision & Sensitivity & F1-score \\
		\hline
  		\multirow{1}{*}{Heart-rate-based} & Ridge Regression & 49.6\% & 53.1\% & 67.0\% & 50.3\% & 54.6\% \\
    \hline
		\multirow{2}{*}{PPG Morphological Features} & Ridge Regression & 53.2\% & 61.4\% & 71.0\% & 54.9\% & 58.9\% \\
		\cline{2-7}
		& Random Forest & 66.7\% & 60.1\% & 74.5\% & 65.3\% & 68.1\% \\	
		\hline
		\multirow{2}{*}{Deep Neural Networks} & ResNet-18 & 67.8\% & 67.5\% & 77.1\% & 67.7\% & 70.4\% \\
		\cline{2-7}
		& ResNet-34 & 66.5\% & 68.4\% & 77.0\% & 66.9\% & 69.7\% \\
	
  \hline
		\multirow{2}{*}{MiniROCKET} & Ridge Regression & 66.2\% & \textbf{69.1\%} & 77.6\% & 66.8\% & 69.4\% \\
\cline{2-7}
		& Random Forest & \textbf{69.3\%} & 68.5\% & \textbf{77.9\%} & \textbf{69.1\%} & \textbf{71.6\%} \\
\hline
	\end{tabular}
		\end{adjustbox}
\label{table:results}
\vspace{-2mm}
\end{table*}
\egroup

\subsection{Baseline Models}
The performance of the proposed method is compared with the following baseline models for the detection of hypertension using PPG signals.

\subsubsection{Heart-rate-based Classifier}
Based on the quasi-periodic nature of the PPG signals, the automatic multiscale-based peak detection (AMPD) algorithm~\cite{scholkmann2012efficient} is used to detect the maximum points. Subsequently, for each PPG signal the average heart rate is calculated. Then, a simple RR classifier is trained for hypertension detection. 
% This simple heart-rate-based classifier is used as the baseline model in this study.

\subsubsection{Classification using PPG Morphological Features}
To extract PPG morphological features, the PPG signals are first segmented using a multi-observation hidden semi-Markov model (HSMM) \cite{hasanzadeh2023multi}. Next, the key points of each PPG pulse, including PPG onset, maximum slope point, systolic peak, dicrotic notch, and diastolic peak, are extracted. Then, BP-related features including heart rate, pulse width, crest time, reflection index, large artery stiffness index (LASI), ratio of PPG pulse areas, and modified normalized pulse volume (mNPV) are derived~\cite{kachuee2016cuffless, hasanzadeh2019blood, wang2023cuffless}.
An RR classifier and an RF classifier are individually trained and evaluated using the extracted features.

\subsubsection{Deep Neural Networks}
Two $1$-D variants of the ResNet-$18$ and ResNet-$34$ deep learning models are trained using the raw PPG signals in an end-to-end manner~\cite{he2016deep}. In these architectures, the $2$-D filters are replaced by $1$-D ones\footnote{\url{https://pypi.org/project/keras-resnet/}}.

\subsection{Training and Validation Setup}

Since the dataset is imbalanced and only a small proportion, around $20\%$, of the training set is labelled as hypertension, all the classifiers were trained using balanced class weights that consider the class frequencies in the input data. For RF classifiers, the model randomly under-samples each bootstrap sample to balance it\footnote{\url{https://pypi.org/project/imbalanced-learn/}}. For deep neural networks, a weighted loss function is used.
Hyperparameters of the models were set using grid-search with respect to the F$1$-score. Adam optimizer was used to train the deep learning models with an initial learning rate of $10^{-3}$, weight decay of $10^{-4}$, and batch size of $32$ for a maximum of $50$ epochs with early-stopping. 
For the MiniROCKET model, the number of kernels is $84$ and the number of output features is  $9,996$.

% The ResNet models are implanted using the Python library \texttt{keras-resnet}\footnote{\url{https://pypi.org/project/keras-resnet/}}.
% The ResNet models are impleneted using the library $keras-resnet$\footnote{\url{https://github.com/navidhasanzadeh/Hypertension_rocket}}. 

% For the model MiniROCKET, the number of kernels was set to the default value proposed in the original paper which is $10,000$.

% All the RF classifiers used in this work are implemented with the \texttt{BalancedRandomForestClassifier} from the library \texttt{imbalanced-learn}\footnote{\url{https://pypi.org/project/imbalanced-learn/}} with $100$ estimators.
 
% Due to the presence of large matrices with dimensions of millions by thousands and the high computational cost in the training phase, the experiments were conducted on a server equipped with an NVIDIA RTX A6000, 400GB of RAM, and an Intel(R) Xeon(R) Platinum 8358 CPU.

\subsection{Results}
In this section, the models implemented in this work are evaluated in terms of sensitivity, precision, and F1-score.

\subsubsection{Classification Performance Analysis}
The performance results in Table \ref{table:results} show that using only the heart-rate feature for hypertension detection leads to a very low sensitivity of $50.3\%$. Although this sensitivity is slightly better than chance level, it is not sufficient on its own for detecting hypertension.

% \subsubsection{PPG Morphological Features}
The extraction of morphological features from PPG signals improves the sensitivity and F1-score for hypertension detection. These features provide more BP-related information than the heart-rate feature alone. Besides, using an RF classifier can discriminate hypertension from normal with significantly better accuracy than RR. The features LASI and reflection index (RI) have the highest Gini indices and importance levels among all the attributes for making the decision trees. These features are related to large artery stiffness and pulse reflection in arteries, respectively.

% \subsubsection{Deep Neural Networks}
Both ResNet models performed better than the models trained on manually-extracted PPG features. This indicates that end-to-end BP-related features extraction from PPG is more robust against PPG signals variations. ResNet-18 obtained a sensitivity of $67.8\%$ for the normal class and $67.5\%$ for detecting hypertension. On average, it performed slightly better than ResNet-34 with an average F1-score of $70.4\%$. 

% \subsubsection{MiniROCKET}
Among all the methods, end-to-end MiniROCKET feature extractor with a balanced RF classifier has achieved the best performance. This method can detect hypertension with an average F1-score of $71.6\%$. The RR classifier with MiniROCKET obtained slightly less average performance but with a sensitivity of $69.1\%$.

The higher performance of the ResNet models and MiniROCKET indicates that there are BP-related features that are not visually observable on PPG pulses and cannot be extracted using manually designed algorithms. Moreover, PPG signals have different shapes among individuals, which makes it difficult to develop a robust manual algorithm for accurate extraction of all the BP-related features.

% It is worth mentioning that although training the models end-to-end, especially MiniROCKET in this work, is computationally expensive, they are still fast enough in the evaluation phase to be implemented on low-end devices, such as smartwatches, for detecting hypertension within seconds.

% \section{Discussion}

\subsubsection{Relationship Between PPG and Hypertension Detection}

The results show that the relationship between high BP and extracted features from a PPG signal is not only limited to heart-rate. By evaluating a range of both manual-based and end-to-end methods on a dataset—where training and testing sets are entirely seperated at the subject level—the findings indicate that PPG signals can be effectively utilized for hypertension detection with a high generalization capability.

% \subsubsection{End-to-end Models Better Performance}

\subsubsection{Impact of the Number of Training Samples}
\begin{figure}[t]
	\centering
	\subfloat{%
		\includegraphics[width=1\linewidth]{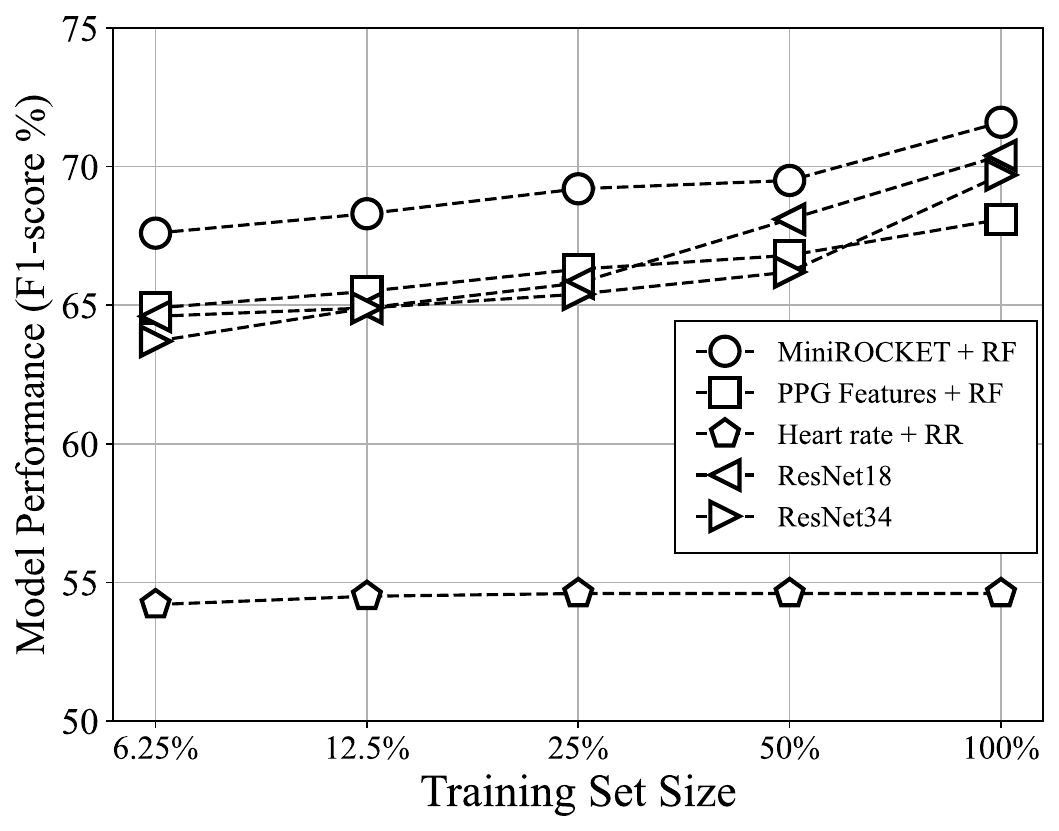}
	}
	\caption{The performance of the models trained on different portions of a training set with one million samples.}
	\vspace{-6mm}
\label{figure:bp}
\end{figure}

In order to study the effect of the size of training set on the models' performance, the models were trained with $6.25\%$, $12.5\%$, $25\%$, and $50\%$ of the training samples. Figure~\ref{figure:bp} illustrates the performance for different models as the training set size increases.
% In this work, the models were trained on one million PPG signals from 3750 subjects. Although the large size of the training samples made training the models computationally challenging,
% it can improve the performance of the models by several percent. For instance,
By using only $50\%$ of the training data, the best F1-score among the models dropped from $71.6\%$ to $69.2\%$. Similarly, using $25\%$ of the training samples resulted in a drop to $65.0\%$. 
% This indicates the direct effect of the training data size on the performance of hypertension detection models. Therefore, we expect that more accurate models can be achieved using larger training datasets, especially with more samples from patients with hypertension.
In all scenarios, MiniROCKET still outperforms all other methods. The trends in this plot indicate that a higher classification accuracy is anticipated by increasing the size of the training dataset, particularly the hypertension data class.
% Figure~\ref{figure:bp} also suggests that there is no improvement for the heart rate-based model when the training sample size increases.

\section{Conclusion}

In this study, the feasibility of hypertension detection using PPG signals is assessed. Utilizing a dataset divided into training, validation, and test sets on a subject basis, the results suggest that the proposed end-to-end method with an RF classifier can achieve an F1-score of $71.6\%$ on the test set. This demonstrates that hypertension detection from PPG signals is capable of generalizing to completely unseen samples. In addition to heart rate, PPG can provide many BP-related informative attributes that can enhance classification performance. The proposed method facilitates the early detection of hypertension using wearable technology.

\addtolength{\textheight}{-12cm}   % This command serves to balance the column lengths

%%%%%%%%%%%%%%%%%%%%%%%%%%%%%%%%%%%%%%%%%%%%%%%%%%%%%%%%%%%%%%%%%%%%%%%%%%%%%%%%

%%%%%%%%%%%%%%%%%%%%%%%%%%%%%%%%%%%%%%%%%%%%%%%%%%%%%%%%%%%%%%%%%%%%%%%%%%%%%%%%

%%%%%%%%%%%%%%%%%%%%%%%%%%%%%%%%%%%%%%%%%%%%%%%%%%%%%%%%%%%%%%%%%%%%%%%%%%%%%%%%

%%%%%%%%%%%%%%%%%%%%%%%%%%%%%%%%%%%%%%%%%%%%%%%%%%%%%%%%%%%%%%%%%%%%%%%%%%%%%%%%

\bibliographystyle{ieeebib}
\bibliography{mybibfile.bib}

\end{document}